# Precise Measurement of the Left-Right Cross Section Asymmetry in $Z$ Boson Production by $e^+e^-$ Collisions

The SLD Collaboration*

*Stanford Linear Accelerator Center*

*Stanford University, Stanford, California, 94309*

## ABSTRACT

We present a precise measurement of the left-right cross section asymmetry ($A_{LR}$) for $Z$ boson production by $e^+e^-$ collisions. The measurement was performed at a center-of-mass energy of 91.26 GeV with the SLD detector at the SLAC Linear Collider (SLC). The luminosity-weighted average polarization of the SLC electron beam was (63.0±1.1)%. Using a sample of 49,392 $Z$ decays, we measure $A_{LR}$ to be 0.1628±0.0071(stat.)±0.0028(syst.) which determines the effective weak mixing angle to be $\sin^2\theta_W^{\rm eff} = 0.2292 \pm 0.0009(\rm stat.) \pm 0.0004(\rm syst.)$.

Submitted to *Physical Review Letters*

This work was supported by Department of Energy contracts: DE-FG02-91ER40676 (BU), DE-FG03-92ER40701 (CIT), DE-FG03-91ER40618 (UCSB), DE-FG02-91ER40672 (Colorado), DE-FG02-91ER40677 (Illinois), DE-FG02-91ER40661 (Indiana), DE-AC03-76SF00098 (LBL), DE-FG02-92ER40715 (Massachusetts), DE-AC02-76ER03069 (MIT), DE-FG06-85ER40224 (Oregon), DE-AC03-76SF00515 (SLAC), DE-FG05-91ER40627 (Tennessee), DE-AC02-76ER00881 (Wisconsin), DE-FG02-92ER40704 (Yale); National Science Foundation grants: PHY-91-13428 (UCSC), PHY-89-21320 (Columbia), PHY-92-04239 (Cincinnati), PHY-88-17930 (Rutgers), PHY-88-19316 (Vanderbilt), PHY-92-03212 (Washington); the UK Science and Engineering Research Council (Brunel and RAL); the Istituto Nazionale di Fisica Nucleare of Italy (Bologna, Ferrara, Frascati, Pisa, Padova, Perugia); and the Japan-US Cooperative Research Project on High Energy Physics (KEK, Nagoya, Tohoku).

*We dedicate this work to the memory of Bill Ash.*

In 1992, the SLD Collaboration performed the first measurement of the left-right cross section asymmetry ($A_{LR}$) in the production of $Z$ bosons by $e^+e^-$ collisions [1]. In this letter, we present a substantially more precise result that is based upon data recorded during the 1993 run of the SLAC Linear Collider (SLC).

The left-right asymmetry is defined as [2],

$$A_{LR} \equiv (\sigma_L - \sigma_R) / (\sigma_L + \sigma_R), \tag{1}$$

where $\sigma_L$ and $\sigma_R$ are the $e^+e^-$ production cross sections for $Z$ bosons at the $Z$ pole energy with left-handed and right-handed electrons, respectively. To leading order, the Standard Model predicts that this quantity depends upon the vector ($v_e$) and axial-vector ($a_e$) couplings of the $Z$ boson to the electron current,

$$A_{LR} = \frac{2v_e a_e}{v_e^2 + a_e^2} = \frac{2\left[1 - 4\sin^2\theta_W^{\rm eff}\right]}{1 + \left[1 - 4\sin^2\theta_W^{\rm eff}\right]^2}, \tag{2}$$

where the effective electroweak mixing parameter is defined [3] as $\sin^2\theta_W^{\rm eff} \equiv (1 - v_e/a_e)/4$. Note that $A_{LR}$ is a sensitive function of $\sin^2\theta_W^{\rm eff}$ and therefore depends upon electroweak radiative corrections including those which involve the top quark and Higgs boson and those arising from new phenomena.

We measure $A_{LR}$ by counting hadronic and $\tau^+\tau^-$ decays of the $Z$ boson for each of the two longitudinal polarization states of the electron beam. The measurement requires knowledge of the absolute beam polarization, but does not require knowledge of the absolute luminosity, detector acceptance, or efficiency [4].

The operation of the SLC with a polarized electron beam has been described previously [5]. The 1993 run of the SLC featured enhanced beam polarization and luminosity. The beam polarization at the SLC source was increased to over 60% by the use of a strained-lattice GaAs photocathode [6] illuminated by a pulsed Ti-Sapphire laser operating at 865 nm [7]. As in 1992, the circular polarization state of each laser pulse (and hence, the helicity of each electron pulse) was chosen randomly.

The maximum luminosity of the collider was increased to $5\times10^{29}$ cm$^2$/sec by the use of flat (elliptical) beams which had transverse aspect ratios of 3/1 [8]. The flat-beam mode of operation precludes the use of the two solenoidal spin rotator magnets (located downstream of the electron damping ring) that were used previously to orient the electron spin direction prior to acceleration in the linac. Therefore, the vertical spin orientation of the beam in the north damping ring is maintained during acceleration and launch into the SLC North Arc. A pair of large amplitude betatron oscillations in the arc is used to adjust the spin direction [9] to achieve longitudinal polarization at the SLC interaction point (IP). The luminosity-weighted mean $e^+e^-$ center-of-mass energy ($E_{cm}$) is measured with precision energy spectrometers [10] to be 91.26±0.02 GeV.

The longitudinal electron beam polarization ($\mathcal{P}_e$) is measured by a Compton scattering polarimeter [11] located 33 m downstream of the IP. After it passes through the IP and before it is deflected by dipole magnets, the electron beam collides with a circularly polarized photon beam produced by a frequency-doubled Nd:YAG laser of wavelength 532 nm. The scattered and unscattered components of the electron beam remain unseparated until they pass through a pair of dipole magnets. The scattered electrons are dispersed horizontally and exit the vacuum system through a thin window. Multichannel Cherenkov and proportional tube detectors measure the momentum spectrum of the scattered electrons in the interval from 17 to 30 GeV/c.

The counting rates in each detector channel are measured for parallel and anti-parallel combinations of the photon and electron beam helicities. The asymmetry formed from these rates is equal to the product $\mathcal{P}_e\mathcal{P}_\gamma A(E)$ where $\mathcal{P}_\gamma$ is the circular polarization of the laser beam at the electron-photon crossing point and $A(E)$ is the theoretical asymmetry function at the accepted energy $E$ of the scattered electrons [12]. For the first 26.9% of the data sample, $\mathcal{P}_\gamma$ was measured to be (97±2)%. For the latter 73.1% of the sample, the laser polarization was maintained at (99.2±0.6)% by continuously monitoring and correcting phase shifts in the laser transport system. The energy scale of the polarimeter is calibrated from measurements of the electron endpoint energy for Compton scattering (17.36 GeV) and the energy at which the asymmetry is zero (25.15 GeV).

Polarimeter data are acquired continually during the operation of the SLC. We obtain $\mathcal{P}_e$ from the observed asymmetry using the measured value of $\mathcal{P}_\gamma$ and the theoretical asymmetry function (including ~1% corrections for detector effects). The measured beam polarization is typically 61-64%. The absolute statistical precision attained in a 3 minute interval is typically $\delta\mathcal{P}_e = 1.0\%$. The systematic uncertainties that affect the polarization measurement are summarized in Table I. The total relative systematic uncertainty is estimated to be $\delta\mathcal{P}_e/\mathcal{P}_e = 1.3\%$.

Due to energy-spread-induced spin diffusion in the SLC arc and imperfect spin orientation, the longitudinal polarization of the electron beam at the IP is typically 95-96% of the polarization in the linac. This result follows from measurements of the arc spin rotation matrix performed with a beam of very small energy spread ($\lesssim 0.1\%$) using the spin rotation solenoids and the Compton polarimeter. These measurements determine the electron polarization in the linac to be (65.7±0.9)%. On several occasions, the beam polarization at end of the linac was directly measured with a diagnostic Møller polarimeter and was found to be (66±3)% [13].

Table I. Systematic uncertainties that affect the $A_{LR}$ measurement.

| Systematic Uncertainty | $\delta\mathcal{P}_e/\mathcal{P}_e$ (%) | $\delta A_{LR}/A_{LR}$ (%) |
|---|---|---|
| Laser Polarization | 1.0 | |
| Detector Linearity | 0.6 | |
| Interchannel Consistency | 0.5 | |
| Analyzing Power Calibration | 0.4 | |
| Electronic Noise Correction | 0.2 | |
| Total Polarimeter Uncertainty | 1.3 | 1.3 |
| Chromaticity Correction ($\xi$) | | 1.1 |
| Corrections in Equation (3) | | 0.1 |
| Total Systematic Uncertainty | | 1.7 |

In our previous Letter [1], we examined a number of effects that could cause the beam polarization measured at the electron-photon crossing point $\mathcal{P}_e$ to differ from the luminosity-weighted beam polarization, $\mathcal{P}_e(1+\xi)$, at the SLC IP. All were found to cause fractional differences $\xi$ that are smaller than 0.001. In 1993, due to very small vertical emittance, the vertical beam size at the IP was limited by third-order chromatic aberrations in the final focus optics. This causes the off-energy electrons to populate the edges of the luminous region. The beam energy and the spin direction at the IP are correlated by the large total spin precession angle in the SLC arc. The on-energy electrons with larger average longitudinal polarization therefore contribute more to the total luminosity and $\xi$ can be non-negligible.

A model based upon the measured energy dependence of the arc spin rotation,

$d\Theta_s/dE = (2.47 \pm 0.03)$ rad/GeV, and the expected dependence of the luminosity on beam energy ($\mathcal{L}(E)$) suggest that $\xi$ is very small ($\xi \lesssim 0.002$) for the Gaussian core ($\Delta E/E \simeq 0.2\%$) of the beam energy distribution, $N(E)$. However, $N(E)$ is observed to have a low-energy tail extending to $\Delta E/E \simeq 1\%$. This small population of low-energy electrons does not contribute to the luminosity but is measured by the polarimeter, leading to a calculated correction factor, $\xi = 0.019 \pm 0.005$. Measurements of $\mathcal{P}_e$ for different settings of an energy-defining collimator agree well with the predictions of the model.

However, we prefer to employ a conservative and essentially model-independent estimate which implicitly includes the energy tail. The correction $\xi$ is rigorously limited to be less than the *difference* between: (1) the observed maximum fractional deviation of $\mathcal{P}_e$ from the polarization in the linac and (2), the minimum fractional deviation of the luminosity-weighted polarization $\mathcal{P}_e(1+\xi)$ from the polarization in the linac. Effect (1) is bounded by our measurements of spin diffusion in the arc to be less than 0.047. Effect (2) is bounded by a calculation using a purely Gaussian energy distribution of narrow width (0.15% RMS), the measured value of $d\Theta_s/dE$, and a chromatically-dominated version of $\mathcal{L}(E)$ to be larger than 0.014. We use the central value and width of the allowed range, 0 to 0.033, to derive the correction factor, $\xi = 0.017 \pm 0.011$, which is applied to our data.

The $e^+e^-$ collisions are measured by the SLD detector which has been described elsewhere [14]. The triggering of the SLD and the selection of $Z$ events are improved versions of the 1992 procedures [1]. The trigger relies on a combination of calorimeter and tracking information, while the event selection is entirely based on the liquid argon calorimeter (LAC) [15]. For each event candidate, energy clusters are reconstructed in the LAC. Selected events are required to contain at least 22 GeV of energy observed in the clusters and to manifest a normalized energy imbalance of less than 0.6 [16]. The left-right asymmetry associated with final state $e^+e^-$ events is expected to be diluted by the t-channel photon exchange subprocess. Therefore, we exclude $e^+e^-$ final states by requiring that each event candidate contain a minimum of 9 clusters (12 clusters if $|\cos\theta|$ is larger than 0.8, where $\theta$ is the angle of the thrust axis [17] with respect to the beam axis).

We estimate that the combined efficiency of the trigger and selection criteria is (93±1)% for hadronic $Z$ decays. Less than 1% of the sample consists of tau pairs. Because muon pair events deposit only small energy in the calorimeter, they are not included in the sample. The residual background in the sample is due primarily to beam-related backgrounds and to $e^+e^-$ final state events. We use our data and a Monte Carlo simulation to estimate the background fraction due to these sources to be $(0.23 \pm 0.10)\%$. The background fraction due to cosmic rays and two-photon processes is (0.02±0.01)%.

A total of 49,392 $Z$ events satisfy the selection criteria. We find that 27,225 ($N_L$) of the events were produced with the left-handed electron beam and 22,167 ($N_R$) were produced with the right-handed beam [18]. The measured left-right cross section asymmetry for $Z$ production is

$$A_m \equiv (N_L - N_R)/(N_L + N_R) = 0.1024 \pm 0.0045.$$

We have verified that the measured asymmetry $A_m$ does not vary significantly as more restrictive criteria (calorimetric and tracking-based) are applied to the sample and that $A_m$ is uniform when binned by the azimuth and polar angle of the thrust axis.

The measured asymmetry $A_m$ is related to $A_{LR}$ by the following expression which incorporates a number of small correction terms in lowest-order approximation,

$$A_{LR} = \frac{A_m}{\langle \mathcal{P}_e \rangle} + \frac{1}{\langle \mathcal{P}_e \rangle}\Bigg[ f_b(A_m - A_b) - A_{\mathcal{L}} + A_m^2 A_{\mathcal{P}} - E_{cm}\frac{\sigma'(E_{cm})}{\sigma(E_{cm})}A_E - A_\varepsilon + A_m\left(1 - \langle \mathcal{P}_e \rangle^2\right)\mathcal{P}_p \Bigg], \tag{3}$$

where $\langle \mathcal{P}_e \rangle$ is the mean luminosity-weighted polarization for the 1993 run; $f_b$ is the background fraction; $\sigma(E)$ is the unpolarized $Z$ cross section at energy $E$; $\sigma'(E)$ is the derivative of the cross section with respect to $E$; $A_b$, $A_{\mathcal{L}}$, $A_{\mathcal{P}}$, $A_E$, and $A_\varepsilon$ are the left-right asymmetries of the residual background, the integrated luminosity, the beam polarization, the center-of-mass energy, and the product of detector acceptance and efficiency, respectively; and $\mathcal{P}_p$ is any longitudinal positron polarization which is assumed to have constant helicity [19].

The luminosity-weighted average polarization $\langle \mathcal{P}_e \rangle$ is estimated from measurements of $\mathcal{P}_e$ made when $Z$ events were recorded,

$$\langle \mathcal{P}_e \rangle = (1+\xi)\cdot\frac{1}{N_Z}\sum_{i=1}^{N_Z}\mathcal{P}_i = (63.0 \pm 1.1)\%, \tag{4}$$

where $N_Z$ is the total number of $Z$ events, and $\mathcal{P}_i$ is the polarization measurement associated in time with the $i^{th}$ event. The error on $\langle \mathcal{P}_e \rangle$ is dominated by the systematic uncertainties on the polarization measurement and the chromaticity correction, $\xi$.

The corrections defined in equation (3) are found to be small. The correction for residual background contamination is moderated by a non-zero left-right background asymmetry ($A_b = 0.031 \pm 0.010$) arising from $e^+e^-$ final states which remain in the sample. The net fractional correction to $A_{LR}$ is $(+0.17 \pm 0.07)\%$. Residual linear polarization of the polarized electron source laser beam can produce a small left-right asymmetry in the electron current ($\lesssim 10^{-3}$). This asymmetry and the left-right asymmetries of all quantities that are correlated with it were reduced by once reversing the spin rotation solenoid at the entrance to the SLC damping ring. The net luminosity asymmetry is estimated from measured asymmetries of the beam current and the rate of radiative Bhabha scattering events observed with a monitor located in the North Final Focus region of the SLC. We determine the left-right luminosity asymmetry to be $A_{\mathcal{L}} = (+3.8 \pm 5.0) \times 10^{-5}$ which leads to a fractional correction of $(-0.037 \pm 0.049)\%$ to $A_{LR}$. A less precise cross check is performed by examining the sample of 125,375 small-angle Bhabha scattering events detected by the luminosity monitoring system (LUM) [20]. Since the left-right cross section asymmetry for small-angle Bhabha scattering is expected to be very small ($\sim -1.5\times10^{-4}\mathcal{P}_e$ in the LUM acceptance), the left-right asymmetry formed from the luminosity Bhabha events is a direct measure of $A_{\mathcal{L}}$. The measured value of $(-32\pm28)\times10^{-4}$ is consistent with the more precisely determined one. The polarization asymmetry is directly measured to be $A_{\mathcal{P}} = (-3.3 \pm 0.1) \times 10^{-3}$, resulting in a fractional correction of $(-0.034\pm0.001)\%$ to $A_{LR}$. The measured left-right beam energy asymmetry of $(+4.4\pm0.1)\times10^{-7}$ arises from the small residual left-right beam current asymmetry due to beam-loading of the accelerator and leads to a fractional correction of $(0.00085\pm0.00002)\%$ to $A_{LR}$. The SLD has a symmetric acceptance in polar angle [4] which implies that the efficiency asymmetry $A_{\varepsilon}$ is negligible. The dominant source of positron polarization [19] is expected to be the Sokolov-Ternov effect in the positron damping ring [21]. Since the polarizing time in the SLC damping rings is about 960 s and the positron storage time is 16.6 ms, the positron polarization emerging from the damping ring is expected to be $1.5 \times 10^{-5}$, leading to a maximum fractional correction of 0.0009% to $A_{LR}$. The corrections listed in equation (3) change $A_{LR}$ by $(+0.10 \pm 0.08)\%$ of the uncorrected value.

Using equation (3), we find the left-right asymmetry to be

$$A_{LR}(91.26 \text{ GeV}) = 0.1628 \pm 0.0071(\text{stat.}) \pm 0.0028(\text{syst.}).$$

The various contributions to the systematic error are summarized in Table I. Correcting this result to account for photon exchange and for electroweak interference which arises from the deviation of the effective $e^+e^-$ center-of-mass energy from the $Z$-pole energy (including the effect of initial-state radiation), we find the pole asymmetry $A^0_{LR}$ and the effective weak mixing angle to be [22],

$$A^0_{LR} = 0.1656 \pm 0.0071(\text{stat.}) \pm 0.0028(\text{syst.})$$
$$\sin^2\theta_W^{\text{eff}} = 0.2292 \pm 0.0009(\text{stat.}) \pm 0.0004(\text{syst.}).$$

We note that this is the most precise single determination of $\sin^2\theta_W^{\text{eff}}$ yet performed. Combining this value of $\sin^2\theta_W^{\text{eff}}$ with our previous measurement at $E_{CM} = 91.55$ GeV [1], we obtain the value, $\sin^2\theta_W^{\text{eff}} = 0.2294 \pm 0.0010$ which corresponds to the pole asymmetry, $A^0_{LR} = 0.1637 \pm 0.0075$. In either form, this result is smaller by 2.3 standard deviations than the average of 25 measurements performed by the LEP Collaborations [23].

We thank the personnel of the SLAC accelerator department and the technical staffs of our collaborating institutions for their outstanding efforts on our behalf. This work was supported by the Department of Energy; the National

Science Foundation; the Istituto Nazionale di Fisica Nucleare of Italy; the Japan-US Cooperative Research Project on High Energy Physics; and the Science and Engineering Research Council of the United Kingdom.

## *The SLD Collaboration

K. Abe,[27] I. Abt,[13] W.W. Ash,[25]† D. Aston,[25] N. Bacchetta,[20]
K.G. Baird,[23] C. Baltay,[31] H.R. Band,[30] M.B. Barakat,[31] G. Baranko,[9]
O. Bardon,[16] T. Barklow,[25] A.O. Bazarko,[10] R. Ben-David,[31]
A.C. Benvenuti,[2] T. Bienz,[25] G.M. Bilei,[21] D. Bisello,[20]
G. Blaylock,[7] J.R. Bogart,[25] T. Bolton,[10] G.R. Bower,[25] J.E. Brau,[19]
M. Breidenbach,[25] W.M. Bugg,[26] D. Burke,[25] T.H. Burnett,[29]
P.N. Burrows,[16] W. Busza,[16] A. Calcaterra,[12] D.O. Caldwell,[6]
D. Calloway,[25] B. Camanzi,[11] M. Carpinelli,[22] R. Cassell,[25]
R. Castaldi,[22][a] A. Castro,[20] M. Cavalli-Sforza,[7] E. Church,[29]
H.O. Cohn,[26] J.A. Coller,[3] V. Cook,[29] R. Cotton,[4] R.F. Cowan,[16]
D.G. Coyne,[7] A. D'Oliveira,[8] C.J.S. Damerell,[24] S. Dasu,[25]
F.J. Decker,[25] R. De Sangro,[12] P. De Simone,[12] S. De Simone,[12]
R. Dell'Orso,[22] Y.C. Du,[26] R. Dubois,[25] J.E. Duboscq,[6]
B.I. Eisenstein,[13] R. Elia,[25] P. Emma,[25] C. Fan,[9] M.J. Fero,[16]
R. Frey,[19] K. Furuno,[19] E.L. Garwin,[25] T. Gillman,[24] G. Gladding,[13]
S. Gonzalez,[16] G.D. Hallewell,[25] E.L. Hart,[26] Y. Hasegawa,[27]
S. Hedges,[4] S.S. Hertzbach,[17] M.D. Hildreth,[25] D.G. Hitlin,[5]
J. Huber,[19] M.E. Huffer,[25] E.W. Hughes,[25] H. Hwang,[19] Y. Iwasaki,[27]
J.M. Izen,[13] P. Jacques,[23] J. Jaros,[25] A.S. Johnson,[3] J.R. Johnson,[30]
R.A. Johnson,[8] T. Junk,[25] R. Kajikawa,[18] M. Kalelkar,[23] I. Karliner,[13]
H. Kawahara,[25] M.H. Kelsey,[5] H.W. Kendall,[16] M.E. King,[25]
R. King,[25] R.R. Kofler,[17] N.M. Krishna,[9] R.S. Kroeger,[26] Y. Kwon,[25]
J.F. Labs,[25] M. Langston,[19] A. Lath,[16] J.A. Lauber,[9] D.W.G. Leith,[25]
T. Limberg,[25] X. Liu,[7] M. Loreti,[20] A. Lu,[6] H.L. Lynch,[25] J. Ma,[29]
G. Mancinelli,[21] S. Manly,[31] G. Mantovani,[21] T.W. Markiewicz,[25]
T. Maruyama,[25] H. Masuda,[25] E. Mazzucato,[11] J.F. McGowan,[13]
A.K. McKemey,[4] B.T. Meadows,[8] R. Messner,[25] P.M. Mockett,[29]
K.C. Moffeit,[25] B. Mours,[25] G. Müller,[25] D. Muller,[25] T. Nagamine,[25]
U. Nauenberg,[9] H. Neal,[25] M. Nussbaum,[8] L.S. Osborne,[16]
R.S. Panvini,[28] H. Park,[19] T.J. Pavel,[25] I. Peruzzi,[12][b] L. Pescara,[20]

M. Piccolo,[12] L. Piemontese,[11] E. Pieroni,[22] K.T. Pitts,[19]
R.J. Plano,[23] R. Prepost,[30] C.Y. Prescott,[25] G.D. Punkar,[25]
J. Quigley,[16] B.N. Ratcliff,[25] T.W. Reeves,[28] P.E. Rensing,[25]
L.S. Rochester,[25] J.E. Rothberg,[29] P.C. Rowson,[10] J.J. Russell,[25]
O.H. Saxton,[25] T. Schalk,[7] R.H. Schindler,[25] U. Schneekloth,[16]
D. Schultz,[25] B.A. Schumm,[15] A. Seiden,[7] S. Sen,[31] M.H. Shaevitz,[10]
J.T. Shank,[3] G. Shapiro,[15] D.J. Sherden,[25] C. Simopoulos,[25]
S.R. Smith,[25] J.A. Snyder,[31] M.D. Sokoloff,[8] P. Stamer,[23]
H. Steiner,[15] R. Steiner,[1] M.G. Strauss,[17] D. Su,[25] F. Suekane,[27]
A. Sugiyama,[18] S. Suzuki,[18] M. Swartz,[25] A. Szumilo,[29]
T. Takahashi,[25] F.E. Taylor,[16] E. Torrence,[16] J.D. Turk,[31] T. Usher,[25]
J. Va'Vra,[25] C. Vannini,[22] E. Vella,[25] J.P. Venuti,[28] P.G. Verdini,[22]
S.R. Wagner,[25] A.P. Waite,[25] S.J. Watts,[4] A.W. Weidemann,[26]
J.S. Whitaker,[3] S.L. White,[26] F.J. Wickens,[24] D.A. Williams,[7]
D.C. Williams,[16] S.H. Williams,[25] S. Willocq,[31] R.J. Wilson,[3]
W.J. Wisniewski,[5] M. Woods,[25] G.B. Word,[23] J. Wyss,[20]
R.K. Yamamoto,[16] J.M. Yamartino,[16] S.J. Yellin,[6] C.C. Young,[25]
H. Yuta,[27] G. Zapalac,[30] R.W. Zdarko,[25] C. Zeitlin,[19] and J. Zhou,[19]

[1] *Adelphi University, Garden City, New York 11530*
[2] *INFN Sezione di Bologna, I-40126 Bologna, Italy*
[3] *Boston University, Boston, Massachusetts 02215*
[4] *Brunel University, Uxbridge, Middlesex UB8 3PH, United Kingdom*
[5] *California Institute of Technology, Pasadena, California 91125*
[6] *University of California at Santa Barbara, Santa Barbara, California 93106*
[7] *University of California at Santa Cruz, Santa Cruz, California 95064*
[8] *University of Cincinnati, Cincinnati, Ohio 45221*
[9] *University of Colorado, Boulder, Colorado 80309*
[10] *Columbia University, New York, New York 10027*
[11] *INFN Sezione di Ferrara and Università di Ferrara, I-44100 Ferrara, Italy*
[12] *INFN Lab. Nazionali di Frascati, I-00044 Frascati, Italy*
[13] *University of Illinois, Urbana, Illinois 61801*
[14] *KEK National Laboratory, Tsukuba-shi, Ibaraki-ken 305 Japan*
[15] *Lawrence Berkeley Laboratory, University of California, Berkeley, California 94720*
[16] *Massachusetts Institute of Technology, Cambridge, Massachusetts 02139*
[17] *University of Massachusetts, Amherst, Massachusetts 01003*
[18] *Nagoya University, Chikusa-ku, Nagoya 464 Japan*
[19] *University of Oregon, Eugene, Oregon 97403*

[20]*INFN Sezione di Padova and Università di Padova, I-35100 Padova, Italy*
[21]*INFN Sezione di Perugia and Università di Perugia, I-06100 Perugia, Italy*
[22]*INFN Sezione di Pisa and Università di Pisa, I-56100 Pisa, Italy*
[23]*Rutgers University, Piscataway, New Jersey 08855*
[24]*Rutherford Appleton Laboratory, Chilton, Didcot, Oxon OX11 0QX United Kingdom*
[25]*Stanford Linear Accelerator Center, Stanford University, Stanford, California 94309*
[26]*University of Tennessee, Knoxville, Tennessee 37996*
[27]*Tohoku University, Sendai 980 Japan*
[28]*Vanderbilt University, Nashville, Tennessee 37235*
[29]*University of Washington, Seattle, Washington 98195*
[30]*University of Wisconsin, Madison, Wisconsin 53706*
[31]*Yale University, New Haven, Connecticut 06511*
†*Deceased*
[a]*Also at the Università di Genova*
[b]*Also at the Università di Perugia*